\documentclass[acmsmall, screen]{acmart}
\usepackage{graphicx}
\usepackage{epstopdf}
\usepackage{tcolorbox}
\usepackage{url}
\usepackage{makecell}
\usepackage{wrapfig}
\usepackage{amsmath,amsfonts}
\usepackage{algpseudocode}
\usepackage{multirow}
\usepackage{array}
\usepackage[ruled,linesnumbered,lined,boxed,commentsnumbered]{algorithm2e}
\usepackage{textcomp}
\usepackage{xcolor}
\usepackage{colortbl}
\definecolor{mygray}{gray}{0.9}
\usepackage{fancybox}
\usepackage{amsthm}
\usepackage{enumitem}
\usepackage{subcaption}
\usepackage{relsize}

\AtBeginDocument{%
  }

\setcopyright{none}
\renewcommand\footnotetextcopyrightpermission[1]{}

\acmISBN{}
\acmDOI{}
\acmPrice{}
\acmYear{2026}
\acmConference[FSE 2026]{}{July 5--9, 2026}{Montreal, Canada}
\acmArticle{1}
\acmMonth{7}

\author{Wei Liu}
\email{weiliu@stu.pku.edu.cn}
\orcid{0009-0005-4603-4227}
\affiliation{
 \institution{Key Lab of High Confidence Software Technologies (PKU), MoE, China}
 \country{China}
}
\affiliation{%
 \institution{School of Computer Science, Peking University
}
 \city{Beijing}
 \country{China}
}

\author{Chao Peng}\authornote{Corresponding authors.}
\email{pengchao.x@bytedance.com}
\orcid{0000-0003-2843-0689}
\affiliation{
 \institution{Bytedance}
 \city{Beijing}
 \country{China}
}

\author{Pengfei Gao}
\email{pengchao.x@bytedance.com}
\orcid{0000-0003-3800-2565}
\affiliation{
 \institution{Bytedance}
 \city{Beijing}
 \country{China}
}

\author{Aofan Liu}
\email{af.liu@stu.pku.edu.cn}
\orcid{0000-0002-6377-1633}
\affiliation{%
 \institution{School of Electronic and Computer Engineering, Peking University}
 \country{China}}

\author{Wei Zhang}\authornotemark[1]
\orcid{0000-0003-1543-0196}
\email{zhangw.sei@pku.edu.cn}
\affiliation{
 \institution{Key Lab of High Confidence Software Technologies (PKU),  MoE, China}
 \country{China}
}
\affiliation{%
 \institution{School of Computer Science, Peking University
}
 \city{Beijing}
 \country{China}
}

\author{Haiyan Zhao}
\orcid{0000-0002-3600-8923}
\email{zhhy.sei@pku.edu.cn}
\affiliation{
 \institution{Key Lab of High Confidence Software Technologies (PKU),  MoE, China}
 \country{China}
}
\affiliation{%
 \institution{School of Computer Science, Peking University
}
 \city{Beijing}
 \country{China}
}

\author{Zhi Jin}\authornotemark[1]
\orcid{0000-0003-1087-226X}
\email{zhijin@pku.edu.cn}
\affiliation{
 \institution{Key Lab of High Confidence Software Technologies (PKU),  MoE, China}
 \country{China}
}
\affiliation{%
 \institution{School of Computer Science, Peking University
}
 \city{Beijing}
 \country{China}
}

\begin{document}

\title{GraphLocator: Graph-guided Causal Reasoning for Issue Localization}


\renewcommand{\shortauthors}{Liu et al.}

\begin{abstract}
The issue localization task aims to identify the locations in a software repository that requires modification given a natural language issue description. 
This task is fundamental yet challenging in automated software engineering due to the semantic gap between issue description and source code implementation.
This gap manifests as two mismatches: (1) \emph{symptom–to-cause mismatches}, where descriptions do not explicitly reveal underlying root causes; (2) \emph{one-to-many mismatches}, where a single issue corresponds to multiple interdependent code entities.
To address these two mismatches, we propose \textsc{GraphLocator}, an LLM-based approach that mitigates symptom–to-cause mismatches through \emph{causal structure discovering} and resolves one-to-many mismatches via \emph{dynamic issue disentangling}.
The key artifact of \textsc{GraphLocator} is the \emph{causal issue graph} (CIG), in which vertices represent discovered sub-issues along with their associated code entities, and edges encode the causal dependencies between them.
The workflow of \textsc{GraphLocator} consists of two phases: \emph{symptom vertices locating} and \emph{dynamic CIG discovering}; it first identifies symptom locations on the repository graph, then dynamically expands the CIG by iteratively reasoning over neighboring vertices, discovering new sub-issues and updating causal dependencies.
Experiments on three real-world Python and Java datasets demonstrates the effectiveness of \textsc{GraphLocator}: (1) Compared with baselines, \textsc{GraphLocator} achieves more accurate localization with average improvements of +19.49\% in function-level recall and +11.89\% in precision with acceptable overhead. 
(2) \textsc{GraphLocator} outperforms baselines on both symptom-to-cause and one-to-many mismatch scenarios, achieving recall improvement of +16.44\% and +19.18\%, precision improvement of +7.78\% and +13.23\%, respectively.
(3) The disentangled causal structure CIG generated by \textsc{GraphLocator} yields the highest relative improvement, resulting in a 28.74\% increase in performance on downstream resolving task.
\end{abstract}

\ccsdesc[500]{Software and its engineering~Software defect analysis}
\ccsdesc[500]{Computing methodologies~Causal reasoning and diagnostics}
\keywords{Issue Localization, Repository Mining, Causal Reasoning}


\maketitle

\section{Introduction}

Software repositories evolve continuously as developers address issues such as feature implementation, bug fixing, and code refactoring~\cite{chen2025locagent, zan2025multiswebench}; a substantial portion of developer effort is devoted to understanding and resolving these issues~\cite{bohme2017bug}.
A critical yet costly initial step in this process is issue localization, which maps a natural language issue description to the specific code snippets that require modification~\cite{xia2024agentless, yang2024swe, chen2025locagent, yu2025orcaloca}.
Despite its importance, automating this task is fundamentally difficult;  even state-of-the-art large language models (LLMs) and frameworks~\cite{wang2024openhands, yang2024swe,xia2024agentless} fail to localize the file to be changed over half of the issues in recent benchmarks~\cite{zan2025multiswebench}. 
This difficulty primarily originates from the semantic gap between the abstract, often ambiguous issue descriptions and the concrete, structured nature of code implementations.

\begin{figure}[h]
    \centering
    \includegraphics[width=0.98\linewidth]{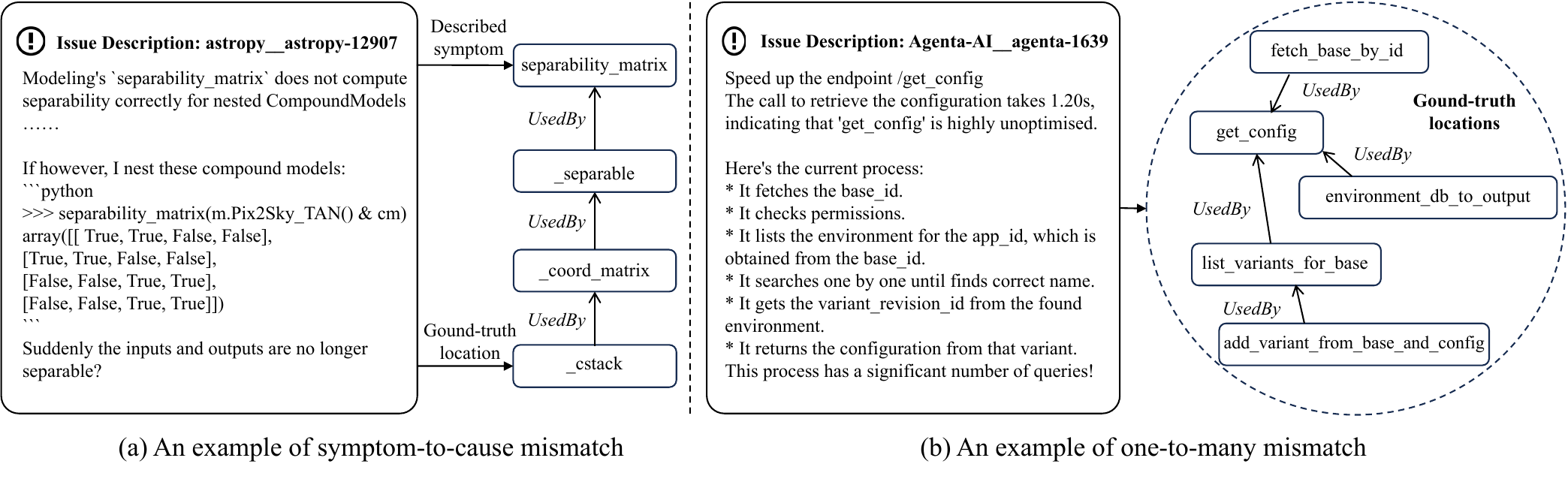}
    \vspace{-3mm}
    \caption{Examples of the two mismatches in issue localization: (a) symptom–to-cause mismatch, where the issue description does not directly expose the underlying root cause; and (b) one-to-many mismatch, where resolving the issue requires modifying multiple interdependent code entities.}
    \label{fig:mismatch}
\end{figure}

To bridge the semantic gap between issue description and source code implementation, 
effective issue localization requires addressing two prevalent types of mismatches: \emph{symptom-to-cause mismatch} and \emph{one-to-many mismatch}.
A symptom–to-cause mismatch arises when issue descriptions report observable symptoms (e.g., ``does not compute separability correctly'' shown in Fig.\ref{fig:mismatch} (a)) instead of the underlying root cause~\cite{chen2025locagent}.
Localizing the root cause therefore requires tracing these implicit, multi-hop causal paths.
A one-to-many mismatch arises when a single high-level issue (e.g., ``speed up the endpoint'' shown in Fig.~\ref{fig:mismatch} (b)) requires coordinated changes across multiple interdependent code entities~\cite{zhou2019understanding}. Effective localization required to identify the full set of related elements, rather than focusing on a single entry point.

\begin{figure}[t]
    \centering
    \includegraphics[width=0.7\linewidth]{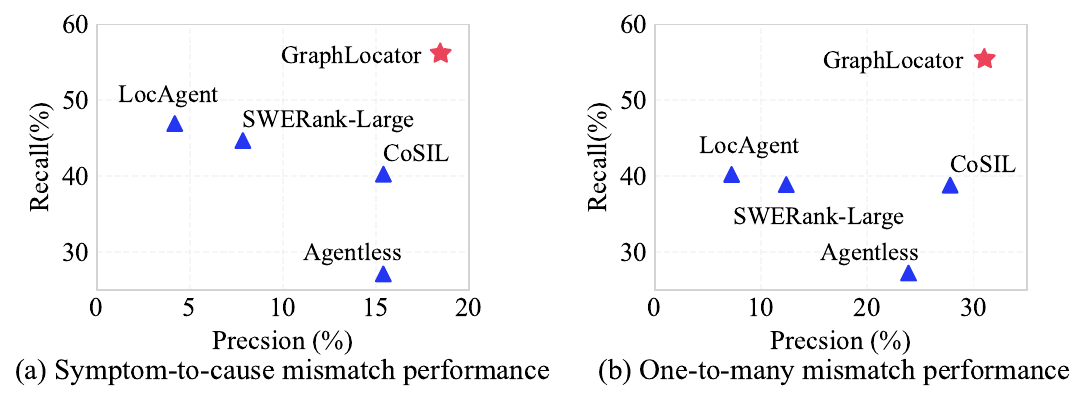}
    \vspace{-3mm}
    \caption{
    Performance of localization approaches in addressing symptom–to-cause (i.e., symptom–to-cause distance > 1) and one-to-many mismatches (i.e., more than one ground-truth function). Results are averaged over issues from SWE-bench Lite, LocBench, and Multi-SWE-bench Java using Claude-3.5.}
    \label{fig:pr}
\end{figure}

Recently, a variety of issue localization approaches have been proposed; however, as illustrated in Fig.\ref{fig:pr}, significant limitations remain in addressing the symptom–to-cause and one-to-many mismatches.
Specifically, existing methods can be largely classified into embedding-based and LLM-based approaches:
(1) \emph{Embedding-based approaches} (e.g., SWERank-Large~\cite{reddy2025swerank}) encode code entities and issue descriptions into embeddings and rank by similarity. They achieve high recall by retrieving broadly related code but fail to capture causal dependencies, resulting in a low localization precision~\cite{wang2023codet5+, sureshcornstack}.
(2) \emph{LLM-based approaches} leverage the reasoning and code understanding capabilities of LLMs, and can be further divided into procedural and agentic workflows. Procedural workflows (e.g., Agentless~\cite{xia2024agentless}) follow a fixed hierarchical process from files to functions. While structured, they typically yield low recall in the presence of symptom–to-cause and one-to-many mismatches due to fixed traversal and the inability to capture cross-hierarchy dependencies. 
Agentic workflows (e.g., LocAgent~\cite{chen2025locagent}, CoSIL~\cite{jiang2025cosil}) allow LLMs to autonomously traverse repository graphs. Although more flexible, they prioritizes superficial relevance instead of underlying causality and struggle to maintain coherent causal chains without explicit contextual guidance~\cite{wei2022chain, luo2025causal}, thus limiting their effectiveness in resolving symptom-to-cause and one-to-many mismatches.

To address these limitations, we propose \textsc{GraphLocator}, an LLM-based approach that addresses symptom–to–cause mismatches through \emph{causal structure discovering} and resolves one-to-many mismatches via \emph{dynamic issue disentangling}. Both of these capabilities are realized via an artifact called the \emph{causal issue graph} (CIG). 
In particular, a CIG is a directed graph in which vertices represent sub-issues and edges encode causal dependencies between them. Each sub-issue is associated with the relevant code snippets it affects.
To construct the CIG, \textsc{GraphLocator} first identifies symptom locations on a \emph{repository dependency fractal structure} (RDFS) using an agentic workflow. 
The RDFS is a heterogeneous graph that includes repository elements of different granularity, such as directories, files, classes, and methods, as vertices, along with the dependency edges connecting them. 
Based on RDFS, \textsc{GraphLocator} then iteratively discovers the causal structure of the issue by performing abductive reasoning over neighboring vertices: each observed neighbor may give rise to a new sub-issue, which is added to the CIG. This process naturally enables dynamic issue disentangling, as the process progressively decouples complex, interdependent issue components while constructing the CIG.

Experiments based on three real-world datasets covering both Python and Java languages demonstrate the effectiveness of \textsc{GraphLocator}.
Compared with embedding- and LLM-based approaches using Claude-3.5 on function-level localization, 
(1) \textsc{GraphLocator} achieves more accurate localization with average absolute improvements of +19.49\% in recall and +11.89\% in precision with acceptable overhead.
(2) \textsc{GraphLocator} outperforms embedding- and LLM-based approaches on both symptom-to-cause mismatches (distance > 1) and one-to-many mismatches, achieving recall improvement of +16.44\% and +19.18\%, precision improvement of +7.78\% and +13.23\%, respectively.
(3) The disentangled structure CIG generated by \textsc{GraphLocator} yields the highest relative improvement, resulting in a 28.74\% increase in performance on the downstream issue-resolving task.

To summarize, our main contributions are:
\begin{itemize}[leftmargin=*]
    \item An LLM-based approach \textsc{GraphLocator} with hybrid workflow that combines agentic exploration of the RDFS with iterative causal structure discovery, effectively tracing root causes and disentangling interdependent code contexts.
    \item A graph-based representation CIG (causal issue graph) that explicitly models causal dependencies between sub-issues, addressing symptom–to-cause mismatches and mitigating one-to-many mismatches by decomposing issues into interdependent sub-issues.
    \item  Experiments on three real-world Python and Java datasets show that \textsc{GraphLocator} significantly improves function-level localization (+19.49\% function-level recall, +11.89\% precision on average), effectively handles symptom-to-cause and one-to-many mismatches, and enhances downstream issue-resolving performance by 28.74\%.
\end{itemize}

\section{Basic Concepts}
In this section, we introduce two concepts used in \textsc{GraphLocator}: the \emph{repository dependency fractal structure} (RDFS) and the \emph{causal issue graph} (CIG).
RDFS provides a structured representation of the repository by modeling code entities and their hierarchical and dependency relations.
CIG captures the causal semantics of a given issue, representing sub-issues grounded to corresponding code entities, along with the inferred causal dependencies among them.

\subsection{Repository Dependency Fractal Structure}\label{sec:rdfs}
The RDFS serves as a unified graph representation that models the context of repositories across various object-oriented programming languages. 
It captures both hierarchical memberships and structural dependencies among elements at different levels of granularity within a repository.
By organizing these elements into a heterogeneous attributed graph, RDFS offers a scalable and flexible way for representing the intricate complexities of code repository.

\begin{definition}[Repository Dependency Fractal Structure] 
A repository dependency fractal structure (RDFS) is a 7-tuple $\mathcal{R} \doteq (V, E, T, C, type, code, layer)$, where: 

\begin{enumerate}[leftmargin=*]
\item $V$ is a set of vertices, with each vertex representing an element in repository and associated with a type in $T_V$;
\item $E \subseteq V \times T_E \times V$ is a set of edges between vertices, with each edge representing a relationship between two elements and associated with a type in $T_E$. 
Given an edge $e = (u,\rho, v) \in E$, $e$ is an \emph{out-edge} of $u$ and an \emph{in-edge} of $v$; 
\item $T = T_V \cup T_E$ is a set of types, where $T_V\!=\!\{\text{\textit{dir, pkg, file, class,}}$ $\text{\textit{interface, enum, field, method,}}$ $\text{\textit{func, global\_var}}\}$ is vertex types, and $T_E\!=\!\{\text{\textit{UsedBy, ImportedBy, ExtendedBy, ImplementedBy}}$, $\text{\textit{HasMember}}\}$ represents edge types; 
\item $C$ is the set of code snippets in repository and each vertex is associated with a code snippet;
\item $type : (V \rightarrow T_V) \cup (E \rightarrow T_E)$ is a function that maps each element in $V \cup E$ to its type; 
\item $code: V \rightarrow C$ is a function that maps each vertex in $V$ to its code snippet in $C$;
\item $layer: T_V \rightarrow \{1, \cdots, L\}$ is a surjective function that maps each vertex type to a layer number.
\end{enumerate}
\end{definition}

The RDFS has three key properties: (1) Edges only connect vertices from one type to those of a type with a layer number that is greater than or equal to its own: $ \forall e=(u, \rho, v) \in E \Rightarrow \textit{layer}(\textit{type}(u)) \leq \textit{layer}(\textit{type}(v))$. 
(2) The only edge type that exists cross layers is \textit{HasMember}: $\forall  e=(u, \rho, v) \in E \wedge layer(type(u)) > layer(type(v)) \Rightarrow \rho = $\textit{ HasMember}.
(3) The layer numbers create a partition of the vertex type set: $\forall i, j \in \{1, \cdots, l\}$, $L_i = \{t \in T_V \mid \text{\textit{layer}}(t) = i\} \neq \emptyset$, $L_i \cap L_j = \emptyset$, and $\bigcup_i L_i = T_V$.

\begin{figure*}[t]
    \centering
    \includegraphics[width=0.98\linewidth]{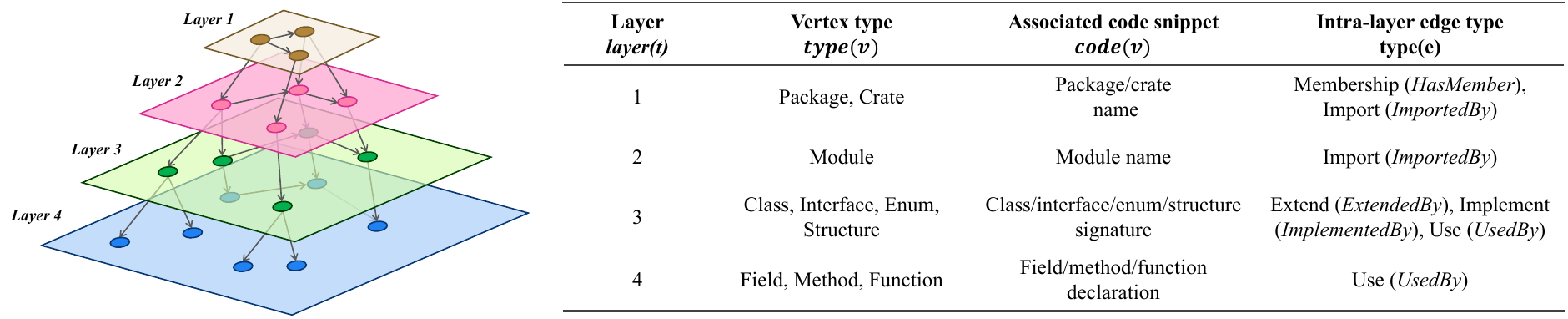}
    \caption{An illustration of RDFS.}
    \label{fig:rdfs}
\end{figure*}

Fig.~\ref{fig:rdfs} gives an illustration of RDFS and its instantiation in programming languages such as Python and Java.
The elements in a code repository can be categorized into $L=4$ distinct layers, each representing a different level of granularity. 
At layer 1, directories and packages serve as logical groupings of related functionality, where directories may contain other directories, forming hierarchical structures connected by \textit{HasMember} relationship.
At layer 2, it consists of files within the repository, which may import other files, represented by the \textit{ImportedBy} edge type.
At layer 3, it includes vertices of types \textit{class}, \textit{interface}, and \textit{enum} make up the third layer, where classes can extend or implement interfaces and reference other classes or enums. Relationships within this level includes \textit{ExtendedBy}, \textit{ImplementedBy} and \textit{UsedBy}.
At layer 4, fields, methods, and functions are represented. The relationships within this layer, including method/function calls and field usage, are denoted by the edge of type \textit{UsedBy}.
Across all layers, the \textit{HasMember} edge type is used to represent hierarchical membership.

To construct the RDFS, we first parse the entire codebase into an abstract syntax tree (AST) using \texttt{tree-sitter}\footnote{\url{https://tree-sitter.github.io}}. 
We then extract code entities by \texttt{tree-sitter}'s pattern-matching language and establish the foundational hierarchical skeleton by adding \textit{HasMember} edges between them. Subsequently, we perform static analysis on the AST to identify semantic dependencies, such as imports, inheritance, and function calls, and encode them as \textit{ImportedBy}, \textit{ExtendedBy}, and \textit{UsedBy} edges. To ensure efficiency, we employ a lazy-loading strategy: the full graph is built incrementally, beginning with the intra-layer \textit{HasMember} skeleton, while the more computationally intensive inter-layer dependency edges are only analyzed and added on-demand during the GraphLocator's traversal. This incremental approach also facilitates updates; for any modified file, we remove its corresponding subgraph and reconstruct it from the updated AST, seamlessly re-integrating it to maintain a consistent graph representation.

\begin{figure}[t]
\centering
\includegraphics[width=0.83\linewidth]{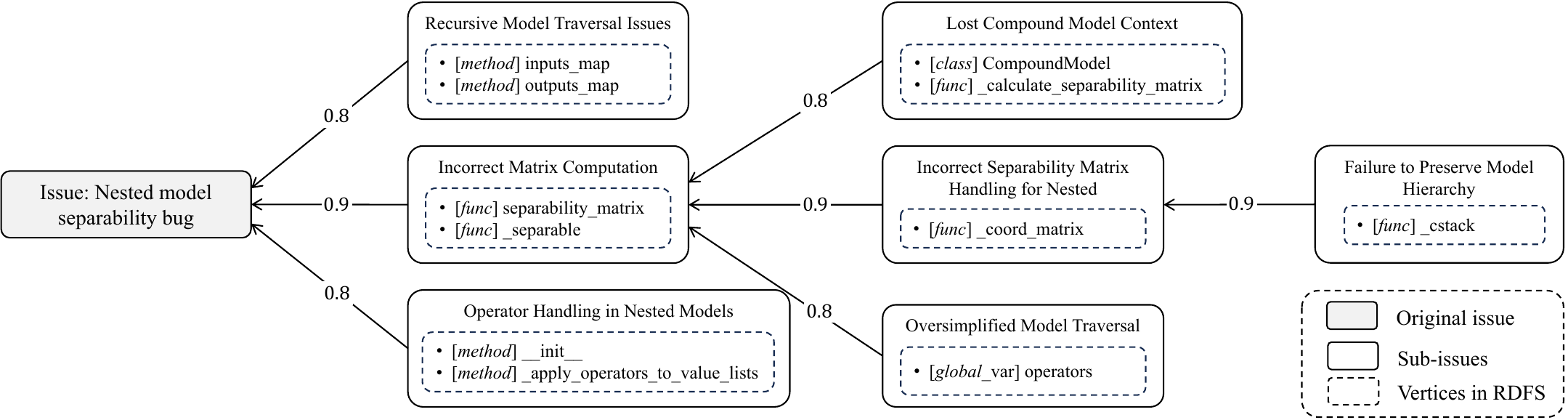}
\vspace{-2mm}
\caption{Illustration of CIG for issue astropy\_\_astropy-12907 in SWE-bench Lite\cite{jimenez2024swebench}. Vertices represent sub-issues grounded to corresponding code entities in RDFS; edges denote probabilistic causal dependencies.}
\vspace{-2mm}
\label{fig:cig}
\end{figure}

\subsection{Causal Issue Graph}

To capture the causal relationships among sub-issues and their associated code entities, we introduce the \emph{Causal Issue Graph (CIG)}. 
Intuitively, a CIG is a directed graph where:
(1) vertices represent sub-issues derived from the issue description and grounded to vertices in RDFS, and 
(2) edges represent potential causal dependencies between these sub-issues. 
Each edge is associated with a probability reflecting the LLM’s estimation of causal probability.

\begin{definition}[Causal Issue Graph]
A \emph{causal issue graph} (CIG) associated with an RDFS $\mathcal{R}$ as the tuple $\mathcal{G}(I, \mathcal{R}) \doteq (\mathcal{X}, \mathcal{Y}, \mathcal{R}, \phi, \psi)$, where:
\begin{enumerate}[leftmargin=*]
    \item $\mathcal{X}$ is the set of sub-issues, each described in natural language;
    \item $\mathcal{Y} \subseteq \mathcal{X} \times \mathcal{X}$ is the set of directed edges representing causal relations between sub-issues;
    \item $\mathcal{R} = (V, E, T, C, \textit{type}, \textit{code}, \textit{layer})$ denotes the corresponding RDFS of the repository;
    \item $\phi : \mathcal{X} \to 2^V$ is a function mapping each sub-issue to its corresponding set of vertices in $\mathcal{R}$;
    \item $\psi : \mathcal{Y} \to [0, 1]$ parametrizes the conditional probability of one sub-issue causing another.
\end{enumerate}
\end{definition}

Conceptually, CIG can be viewed as a domain-specific form of structural causal model (SCM)~\cite{pearl2010causal, pearl2009causal, pearl2009causality}, 
but grounded in repositories: 
vertices(sub-issues) correspond to SCM's causal variables, edges encode causal dependencies, $\psi$ parametrizes conditional effects, and the mapping $\phi$ ties vertices to concrete code. 
Unlike classical SCMs, however, the CIG does not attempt to identify statistically validated causal effects. Instead, it provides a structured representation of hypothesized cause–effect relations among sub-issue to guide LLM-driven abductive reasoning, helping the model to trace multi-hop causal paths and disentangle interdependent sub-issues. 

Fig.\ref{fig:cig} shows a CIG generated by Claude-3.5 for issue \texttt{astropy\_\_astropy-12907} in SWE-bench Lite\cite{jimenez2024swebench}.
The CIG begins from symptom-level functions (e.g., \textit{separability\_matrix} mentioned in the issue description) and, rather than jumping directly to a candidate root cause, incrementally introduces intermediate sub-issues (e.g., incorrect handling of nested separability matrices) by tracing dependencies in the RDFS. These intermediate nodes act as explanatory bridges, linking observable symptoms to the underlying defect. By explicitly modeling them, the CIG enables graph-guided causal reasoning: the LLM can disentangle sub-issues, trace how high-level symptoms propagate through repository dependencies, and finally converge on the true root causes (e.g., \texttt{\_cstack}).
As illustrated in Fig.~\ref{fig:cig}, the CIG resembles a reversed tree that aggregates multiple causal paths. 
However, we use the term \emph{graph} rather than \emph{tree} to maintain generality, since causal paths may share intermediate nodes or intersect. Such overlaps introduce non-tree structures that a tree representation cannot adequately capture.

\section{GraphLocator}

\begin{figure*}[t]
    \centering
    \includegraphics[width=\linewidth]{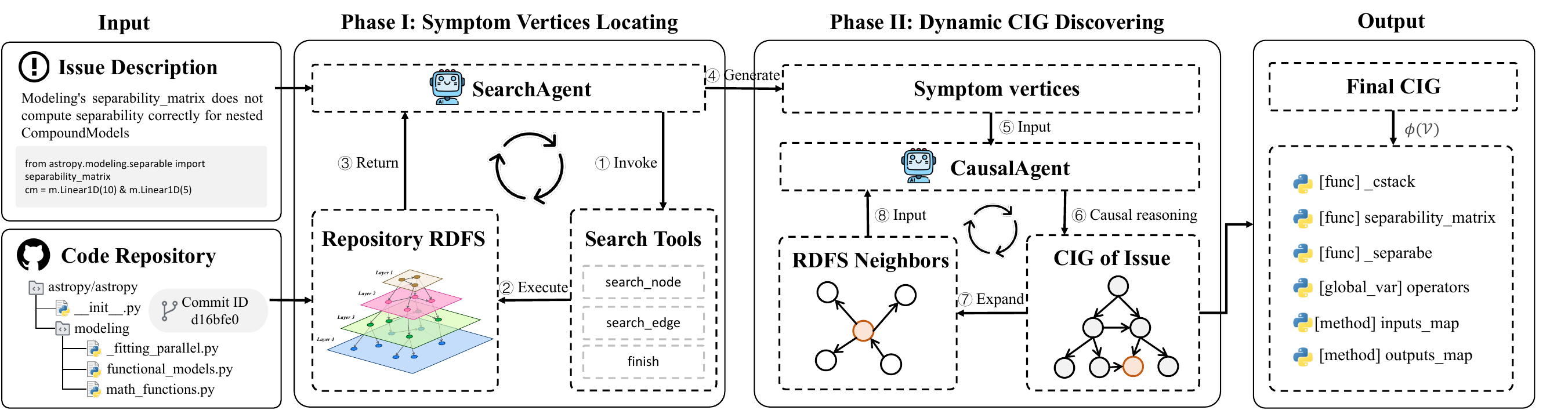}
    \caption{Overview of \textsc{GraphLocator} for issue localization.}
    \vspace{-3mm}
    \label{fig:graphlocator}
\end{figure*}

In this section, we present \textsc{GraphLocator}, a graph-based issue localization framework.
As shown in Fig.~\ref{fig:graphlocator}, 
\textsc{GraphLocator} takes as input an issue description as well as the corresponding code repository, outputs a set of code entities that required modification.
Specifically, \textsc{GraphLocator} consists of two consecutive phases: symptom vertices locating and dynamic CIG discovering.

\begin{itemize}[leftmargin=*]

\item In the \emph{symptom vertices locating} phase (Section~\ref{sec:seed_locations_selecting}), 
\textsc{GraphLocator} aims to identify vertices in the RDFS that correspond to the symptoms described in the issue report by iteratively invoking graph-executable search tools.
\item In the \emph{dynamic CIG discovering} phase (Section~\ref{sec:recursive_subgraph_reasoning}),  
starting from symptom vertices, \textsc{GraphLocator} incrementally constructs the CIG by iteratively traversing neighboring RDFS vertices, discovering new sub-issues, and grounding them to relevant code entities. This process supports multi-hop causal reasoning and disentangles interdependent components, addressing both symptom-to-cause and one-to-many mismatches.
\end{itemize}

Due to space constraints, the prompt templates used in these two phases are omitted here but are provided in the artifact accessible via the data availability link.

\subsection{Phase I: Symptom Vertices Locating}\label{sec:seed_locations_selecting}

The symptom vertices locating phase aims to identify a set of symptom vertices $S$ in the RDFS that directly correspond to issue description.  To achieve this, \textsc{GraphLocator} employs an LLM-driven agent, \emph{SearchAgent}, which iteratively invokes graph-executable search tools to autonomously identify vertices that are semantically aligned with the issue description.

Tab.\ref{tab:search_tool_design} summarizes the search tool set of SearchAgent. 
This set is designed to handle issue descriptions, which are often unstructured, ambiguous, and vary in granularity. Such characteristics pose two specific challenges: (1) descriptions span multiple abstraction levels, and (2) they may embed implicit or explicit structural constraints. 
To address these challenges, \textsc{GraphLocator} enhances the expressiveness of the search tools compared to prior work\cite{chen2025locagent, jiang2025cosil} in three key ways: (1) enabling layer-aware queries by incorporating vertex types, (2) supporting wildcard-based partial matching, and (3) introducing a structure-aware search tool, \textit{search\_edge}, to capture dependencies between entities.
Specifically, the tools include:

\begin{itemize}[leftmargin=*]
    \item \textit{search\_vertex}: This tool retrieves vertices from the RDFS that match the specified constraints, namely $vertex\_name$ and $vertex\_type$. To accommodate the inherent ambiguity and variability of natural language issue description, wildcards (\texttt{*}) can be used for either field, for example \textit{search\_vertex(vertex\_name, \texttt{*})}, enabling partial specification when full details are unavailable. In cases where no exact match exists, the tool performs a fuzzy search based on string edit distance~\cite{ristad2002learning, masek1980faster}, providing robustness against minor variations in naming or typographical errors. The top-k candidates retrieved are then filtered using prompt-based semantic reasoning by the SearchAgent, ensuring that only vertices most semantically aligned with the issue description are retained. During this filtering, each vertex $v$ is serialized in the prompt by concatenating its file path, line number, and $code(v)$ to facilitate judgment by the SearchAgent.

    \item \textit{search\_edge}: This tool identifies edges between vertices that satisfy specified relational constraints. For example, \textit{search\_edge(class, CompoundModel, HasMember, method, \_\_init\_\_)} retrieves the \textit{CompoundModel} class vertex along with its constructor method \textit{\_\_init\_\_}. By supporting wildcards similar to \textit{search\_vertex}, the tool allows flexible specification of edge endpoints or types. For example, \textit{search\_edge(\texttt{*}, \texttt{*}, UsedBy, func, separability\_matrix)} retrieves all code entities that invoke the function \textit{separability\_matrix}. This tool enables the SearchAgent to capture explicit and implicit dependencies in issue descriptions, thus facilitating accurate localization.
    \item \textit{finish}: This tool signals the end of the current phase. When invoked, it returns the collected vertices from the RDFS corresponding to the issue description, allowing the system to proceed to subsequent CIG discovering phase.
\end{itemize}

\begin{table}[t]
    \centering
    \caption{Specification of search tools. Wildcard symbols (\texttt{*}) are allowed for partial argument matching.}
    \vspace{-3mm}
    \scalebox{0.75}{
    \begin{tabular}{m{2cm}m{2.5cm}m{8cm}}
    \toprule
    \textbf{Tool Name} & \textbf{Arguments} & \textbf{Description} \\
    \hline
    search\_vertex & vertex\_name \quad
    vertex\_type & Retrieve a vertex in RDFS by matching its name and type. \\
    \hline
    search\_edge & src\_vertex\_type src\_vertex\_name edge\_type trg\_vertex\_type trg\_vertex\_name & Query an edge between two vertices, specified by type and name, with a given relation. \\
    \hline
    finish & - & Finish and return the final results. \\
    \bottomrule
    \end{tabular}}
    \label{tab:search_tool_design}
\end{table}

\subsection{Phase II: Dynamic CIG Discovering}\label{sec:recursive_subgraph_reasoning}

Given a repository RDFS $\mathcal{R}$ and an issue description $I$, \textsc{GraphLocator} incrementally constructs a CIG to unfold the underlying structure of the issue. The construction expands the CIG in a priority-driven manner, while applying graph-guided abductive reasoning to incorporate newly discovered vertices. Details are given in Alg.~\ref{alg:codescm}.

\paragraph{Priority-driven graph expansion}
This step determines the order in which sub-issues are expanded during CIG construction. 
Agent-based approaches often suffer from incoherent multi-step reasoning, as the LLM may simultaneously handle multiple loosely connected sub-issues and lose causal focus~\cite{luo2025causal,wei2022chain}. 
To mitigate this, we adopt a procedural expansion strategy in which each iteration isolates and expands a single sub-issue.  
Given a CIG $\mathcal{G}^t$ at step $t$, the task is to select the most promising sub-issue $x$ for expansion. 
Intuitively, $x$ should be the vertex with the highest potential to causally influence other issues. 
We formalize this intuition using the priority score $\Psi(x)$:  
\begin{equation*}
    \Psi(x) = 1 - \prod_{(x,y) \in \mathcal{Y}} \bigl(1 - \psi(x,y)\bigr)
\end{equation*}
where $\psi(x,y)$ denotes the LLM-estimated probability that $x$ causally leads to $y$. 
At each step, the sub-issue with the largest $\Psi(x)$ is dequeued and expanded (line 7 in Alg.~\ref{alg:codescm}), while newly discovered sub-issues are inserted into the priority queue with their own $\Psi(\cdot)$ values (line 10 in Alg.~\ref{alg:codescm}). 
This design ensures that the expansion process remains both focused and causally coherent, progressively prioritizing sub-issues most likely to reveal underlying root causes.  

\paragraph{Graph-guided Abductive Reasoning}
This step determines how sub-issues are expanded during CIG construction by guiding the LLM's causal reasoning with structured context. 
At step $t$, given an issue description $I$, the CIG from the previous iteration $\mathcal{G}^{t-1}$, and the sub-issue $x$ to be expanded, we first identify the direct causal candidates of $x$ as the neighboring code vertices of $\phi(x)$ that have not yet been visited. 
These vertices are collected as newly observed nodes $\mathcal{O}^t$  (line 8 in Alg.~\ref{alg:codescm}).

To perform abductive reasoning, we construct a structured prompt for the LLM consisting of: (1) the issue description $I$, (2) the serialized CIG $\mathcal{G}^{t-1}$ in Mermaid format~\cite{sveidqvist2014mermaid}, (3) the target sub-issue $x$, and (4) the list of newly observed code vertices $\mathcal{O}^t$. 
Mermaid is used since it provides a concise, text-based representation of graph data with explicit causal edges (e.g., A --> B), which LLMs parse effectively than Json or adjacent format, allowing robust reasoning over causal relationships~\cite{chen2025accurate, li2025reasongraph}.
Given this structured context, LLM evaluates whether each candidate in $\mathcal{O}^t$ constitutes a plausible cause or intermediary for $x$ and updates the CIG accordingly  (line 9 in Alg.~\ref{alg:codescm}). 
By iteratively applying this procedure, \textsc{GraphLocator} incorporates causally coherent and decoupled sub-issues, incrementally constructing a causal chain from observed symptoms to root causes.

\begin{figure}[t]
\centering
\scalebox{0.88}{%
\begin{minipage}{1.1\linewidth} 
\begin{algorithm}[H]
\small
\SetKwInOut{Input}{Input}
\SetKwInOut{Output}{Output}
\caption{Phase II - Dynamic CIG Discovering}
\label{alg:codescm}
\Input{Repository RDFS $\mathcal{R} = (V, E, T, C, \textit{type}, \textit{code}, \textit{layer})$, issue description $I$, symptom vertices $S \subseteq V$, and maximum number of turns \textit{max\_turn}}
\Output{CIG $\mathcal{G} = (\mathcal{X},\mathcal{Y},\mathcal{R},\phi,\psi)$}

Initialize $t \gets 0$, visited set $\mathcal{V} \gets \emptyset$, 
observed set $\mathcal{O}^0 \gets S$, and represent issue $I$ as a vertex $x_I$\;
$\mathcal{G}^0 \gets \textsc{CausalAgent}(I, \emptyset, x_I,  \mathcal{O}^0)$ \;
Initialize priority queue $Q$ with $\{(x, \Psi(x)) \mid x \in \mathcal{X}^0, x \neq x_I\}$ sort descending with $\Psi(x)$\;
$\mathcal{V} \gets \bigcup_{x \in \mathcal{X}^0} \phi(x)$ \;
\While{$Q$ is not empty and $t < max\_turn$}{
    $t \gets t+1$ \;
    $x \gets Q.\text{pop}()$\;
    $\mathcal{O}^t \gets \{ u \in V \mid \exists v \in \phi(x) \backslash \mathcal{V}, (u,\rho,v)\in E \lor (v,\rho,u)\in E \}$\;
    $\mathcal{G}^t \gets \textsc{CausalAgent}(I, \mathcal{G}^{t-1}, x, \mathcal{O}^t)$ \;
    Insert $\mathcal{X}^t \backslash \mathcal{X}^{t-1}$ into $Q$ with priority $\Psi(x)$\;
    $\mathcal{V} \gets (\bigcup_{x \in \mathcal{X}^t} \phi(x)) \cup \mathcal{V}$\;
}
\Return{Final CIG $\mathcal{G}^t$}\;
\end{algorithm}
\end{minipage}
}
\end{figure}

\section{Experimental Setups}
To evaluate the performance of \textsc{GraphLocator} on issue localization tasks, we have formulated the following four research questions (RQs):

\begin{itemize}[leftmargin=*]
\item \textbf{RQ1 (Effectiveness):} 
How effective is \textsc{GraphLocator} in localization compared to baselines?
\item \textbf{RQ2 (Generalizability):} How does \textsc{GraphLocator} perform across tasks of varying complexity, particularly in scenarios involving symptom-to-cause and one-to-many mismatches?
\item \textbf{RQ3 (Ablation):} How does each component of \textsc{GraphLocator} contribute to its performance?
\item \textbf{RQ4 (Cost):} What is the computational cost of \textsc{GraphLocator} compared with other approaches?
\end{itemize}

\subsection{Datasets}

To comprehensively evaluate issue localization performance, we employ three datasets covering two programming languages (i.e., Python and Java).
The three datasets consist of issues from GitHub repositories, where each issue includes (1) an issue description, (2) the repository commit version, and (3) the corresponding diff-based fix patch.

\begin{itemize}[leftmargin=*]
\item \textbf{SWE-bench Lite (Python)~\cite{jimenez2024swebench}:} 
A widely used benchmark for issue localization and resolving~\cite{reddy2025swerank, chen2025locagent, jiang2025cosil, yang2025enhancing, xia2024agentless}. It comprises 300 issues from 11 large-scale Python projects, selected heuristically from the full 2294-issue SWE-bench. This 300-issue subset is designed to reduce evaluation costs while maintaining benchmark quality. To ensure clean, text-based evaluation, tasks containing non-textual elements (e.g., images and external hyperlinks) are excluded.

\item \textbf{LocBench (Python)~\cite{chen2025locagent}:} 
A recent dataset specifically designed for code localization task, including 560 issue instances from 164 Python repositories.
Unlike SWE-bench Lite, whose issues are mostly bug report, LocBench covers a broader range of issue types and more balanced distribution, including feature requests, security issues, performance problems, and also bug reports.
As the repository \texttt{NCSU-High-Powered-Rocketry-Club/AirbrakesV2} was no longer accessible as of June 25, 2025, its issue instance is excluded, resulting in a final dataset of 559 issues.

\item \textbf{Multi-SWE-bench (Java)~\cite{zan2025multiswebench}:} This dataset consists of 128 issue instances from 9 open-sourced Java projects, supporting both issue localization and resolving tasks.
The dataset is constructed through a systematic pipeline that ensures quality and reliability, including selection of high-quality GitHub repositories based on stars and runnability, and manual verification via dual annotation and cross-review to produce human-verified ground truth.
\end{itemize}

\paragraph{Extraction of Ground-truth Locations.}
Following prior work~\cite{jiang2025cosil, chen2025locagent}, ground-truth locations are extracted from diff-based fix patches at three levels of granularity. 
However, the ground-truth extracted in these works contains missing locations for part of issues.
To remedy this, the ground truth is re-extracted based on RDFS:
At the \emph{file level}, the paths of all modified files are recorded. At the \emph{module level}, the enclosing structural unit in RDFS (e.g., class, enum, or interface) that contains the modified lines is identified. At the \emph{function level}, the function or method in RDFS that directly contains the modified lines is extracted. This hierarchical definition ensures consistency with existing benchmarks while providing fine-grained supervision for evaluation.

\subsection{Evaluation Metrics}
We evaluate issue localization performance at three levels of granularity: file, module, and function.
At each level, we adopt four complementary metrics, Success Location, Recall, Precision, and F1-score.
Formally, let $\mathcal{I}$ denote the set of issue instances, $L_i$ the ground-truth locations for issue $i \in \mathcal{I}$, and $L'_i$ the predicted locations at the file/module/function level. The metrics are defined as follows:

\begin{itemize}[leftmargin=*]
    \item \textbf{Success Location (SL)}~\cite{xia2024agentless, chen2025locagent, yu2025orcaloca} measures if the predicted set fully covers the ground truth.
    \begin{equation*}
        SL = \frac{1}{|\mathcal{I}|} \sum_{i \in \mathcal{I}}\mathbb{1}_{L_i \subseteq L'_i}
    \end{equation*}
    \item \textbf{Recall (REC)} is the proportion of ground-truth locations captured by localization approaches.
    \begin{equation*}
        REC = \frac{1}{|\mathcal{I}|} \sum_{i \in \mathcal{I}}\frac{L_i \cap L'_i}{|L_i|}
    \end{equation*}
    \item \textbf{Precision (PRE)} penalizes overprediction by the fraction of predictions that are correct.
    \begin{equation*}
        PRE = \frac{1}{|\mathcal{I}|} \sum_{i \in \mathcal{I}}\frac{L_i \cap L'_i}{|L'_i|}
    \end{equation*}
    \item \textbf{F1-Score (F1)} provides a harmonic mean of precision and recall, offering a balanced measure that accounts for both under- and over-localization.  
    \begin{equation*}
        F1 = \frac{1}{|\mathcal{I}|} \sum_{i \in \mathcal{I}} \frac{2 \cdot |L_i \cap L'_i|}{|L_i| + |L'_i|}
    \end{equation*}
\end{itemize}

\subsection{Methods for Comparison}
To ensure a comprehensive evaluation, \textsc{GraphLocator} is compared with representative baselines from embedding- and LLM-based approaches.
For each of these two categories, baselines are chosen based on performance: if one approach clearly outperforms another in prior work, only the superior one is included. 
The selected baselines are:
\begin{itemize}[leftmargin=*]
    \item \textbf{SWERank-Small/Large}~\cite{reddy2025swerank}: 
    An embedding-based approach for issue localization that follows a retrieve-then-rerank workflow. 
    Specifically, SWERank-Small combines SWERankEmbed-Small (137M) for retrieval with SWERankLLM-Small (7B) for reranking, 
    while SWERank-Large combines SWERankEmbed-Large (7B) with SWERankLLM-Large (32B). The top-10 reranked results are used as the final localization outputs.
    \item \textbf{Agentless}~\cite{xia2024agentless}: An LLM-based approach with procedural workflow, which follows a hierarchical localization process: it first identifies suspicious files, then narrows down to classes or functions. At each level, the LLM is used to rank and select the top-$N$ most likely locations.
    \item \textbf{LocAgent}~\cite{chen2025locagent}: An LLM-based approach with agentic workflow, which first parses the codebase to construct a graph representation and then builds sparse indexes. Based on these indexes, LocAgent performs agent-guided searches with tool-assisted retrieval.
    \item \textbf{CoSIL}~\cite{jiang2025cosil}: An LLM-based approach with agentic workflow, which uses module and function call graphs to identify suspicious code not explicitly mentioned in the issue. It performs iterative searches over the call graph, guided by a pruning mechanism that restricts exploration to relevant paths and effectively manages contextual information.
\end{itemize}

Note that SWERank and LocAgent are specific to Python: SWERank is fine-tuned on Python repositories, and LocAgent employs a Python-specific backbone graph. Consequently, their evaluation is restricted to SWE-Bench Lite and LocBench, excluding Multi-SWE-bench Java.

\subsection{Implementation Details}

\paragraph{Base LLMs.}
For \textsc{GraphLocator} and LLM-based baselines (i.e., Agentless, LocAgent, and CoSIL), we implement using two representative LLMs: gpt-4o-2024-1120 and claude-3-5-sonnet-20241022, short for GPT-4o and Claude-3.5, respectively. To ensure a fair comparison, both the two models are configured with a greedy decoding (i.e., temperature of 0) to minimize randomness.

\emph{Implementation of Localization Approaches.}
GraphLocator’s RDFS construction is described in Section~\ref{sec:rdfs}. 
For localization, Phase I is limited to 5 search iterations, and Phase II allows up to 20 turns ($max\_turn=20$). 
Baseline approaches (i.e., SWERank, Agentless, LocAgent, and CoSIL) are implemented using their official GitHub repositories with default hyperparameters.
To evaluate Java repositories, the Multi-SWE-Bench version of Agentless\footnote{\url{https://github.com/multi-swe-bench/MagentLess}} is used, and CoSIL is adapted to Java by leveraging this Java-specific code structure generation.

\emph{Hardware Environments} All experiments are conducted on a cluster equipped with 64 cores Intel(R) Xeon(R) Platinum 8457C processor and Debian 10 Linux 5.4.

\section{Experimental Results and Analysis}

\subsection{RQ1: Effectiveness}

Tab.~\ref{tab:effectiveness} shows the effectiveness of \textsc{GraphLocator} and baseline approaches across the three datasets: SWE-bench Lite, LocBench, and Multi-SWE-bench Java. 
Based on these results, several key findings can be summarized as follows:

\begin{table*}[t]
    \centering
    \caption{Effectiveness on the Python (i.e., SWE-bench Lite and LocBench) and Java (i.e., Multi-SWE-bench Java) datasets. Results show file-, module-, and function-level performance. All values are in percent(\%). Bold numbers indicate the best results per base model. Underlined numbers indicate the best results across models and approaches for the same metrics.}
    \vspace{-3mm}
     \scalebox{0.75}{
    \begin{tabular}{llcccc|cccc|cccc}
    \toprule
    \multirow{2}{*}{\textbf{Base Model}} & \multirow{2}{*}{\textbf{Approach}} & \multicolumn{4}{c|}{\textbf{File level}} & \multicolumn{4}{c|}{\textbf{Module level}} & \multicolumn{4}{c}{\textbf{Function level}} \\
    \cmidrule{3-14} 
    & & \textbf{SL} & \textbf{REC} & \textbf{PRE} & \textbf{F1} & \textbf{SL} & \textbf{REC} & \textbf{PRE} & \textbf{F1} & \textbf{SL} & \textbf{REC} & \textbf{PRE} & \textbf{F1} \\
    \hline
    \rowcolor{mygray}\multicolumn{14}{c}{\textbf{SWE-bench Lite}} \\
    \multirow{2}{*}{SWERank}  & Small & 80.67 &  80.67 &  27.84 &  38.55 &  75.11 &  76.00 &  22.48 &  32.96 &  51.26 &  52.44 &  6.06 &  10.73 \\
     & Large  & \textbf{84.33} &  \textbf{84.33} &  \textbf{29.85} &  \textbf{41.28} &  \textbf{78.67} &  \textbf{79.78} &  \textbf{26.27} &  \textbf{36.72} &  \textbf{55.23} &  \textbf{56.50} &  \textbf{6.57} &  \textbf{11.61} \\
    \hline
    \multirow{4}{*}{GPT-4o} & Agentless & 77.00 &  77.00 &  25.67 &  38.50 &  66.67 &  67.33 &  28.70 &  38.08 &  53.43 &  55.72 &  14.44 &  20.70 \\ 
     & LocAgent & 82.67 &  82.67 &  26.50 &  37.95 &  64.89 &  65.63 &  10.64 &  17.15 &  51.99 &  54.48 &  7.24 &  12.28 \\ 
     & CoSIL & 61.67 &  61.67 &  12.49 &  20.77 &  60.44 &  61.41 &  29.79 &  37.81 &  46.21 &  48.35 &  11.37 &  18.05 \\ 
     & GraphLocator & \textbf{83.67} &  \textbf{83.67} &  \underline{\textbf{48.90}} &  \underline{\textbf{56.76}} &  \textbf{70.67} &  \textbf{71.78} &  \underline{\textbf{41.01}} &  \underline{\textbf{47.83}} &  \textbf{60.65} &  \textbf{63.18} &  \underline{\textbf{20.82}} &  \textbf{27.29} \\ 
    \hline
    \multirow{4}{*}{Claude-3.5} & Agentless & 80.00 &  80.00 &  26.67 &  40.00 &  71.56 &  72.52 &  37.15 &  46.28 &  37.91 &  40.01 &  18.75 &  23.27 \\
     & LocAgent & 84.33 &  84.33 &  29.79 &  41.15 &  72.44 &  73.93 &  6.55 &  10.87 &  58.48 &  61.94 &  5.01 &  8.60 \\
     & CoSIL & 82.33 &  82.33 &  16.91 &  28.03 &  74.67 &  75.63 &  34.31 &  44.86 &  59.21 &  62.27 &  15.94 &  24.69 \\
     & GraphLocator & \underline{\textbf{89.33}} &  \underline{\textbf{89.33}} &  \textbf{45.81} &  \textbf{55.17} &  \underline{\textbf{80.44}} &  \underline{\textbf{81.78}} &  \textbf{38.63} &  \textbf{46.82} &  \underline{\textbf{70.76}} &  \underline{\textbf{73.23}} &  \textbf{20.51} &  \underline{\textbf{27.73}} \\
    \bottomrule
    \rowcolor{mygray}\multicolumn{14}{c}{\textbf{LocBench}} \\
    \multirow{2}{*}{SWERank}  & Small & 75.85 &  81.17 &  30.65 &  40.51 &  68.17 &  73.61 &  \textbf{26.50} &  35.29 &  37.52 &  43.47 &  7.63 &  12.03 \\
     & Large & \textbf{78.35} &  \textbf{83.19} &  \textbf{33.29} &  \textbf{43.18} &  \textbf{70.43} &  \textbf{75.65} &  26.31 &  \textbf{35.79} &  \textbf{38.42} &  \textbf{44.98} &  \textbf{7.88} &  \textbf{12.45} \\
    \hline
    \multirow{4}{*}{GPT-4o} & Agentless & 67.98 &  72.95 &  28.50 &  39.67 &  59.65 &  65.10 &  33.87 &  40.82 &  44.88 &  53.90 &  20.06 &  25.07  \\ 
     & LocAgent & 74.06 &  78.86 &  28.21 &  38.54 &  59.15 &  65.02 &  11.14 &  17.38 &  43.99 &  53.96 &  9.82 &  14.53 \\ 
    & CoSIL & 56.71 &  60.75 &  15.83 &  23.88 &  47.87 &  52.89 &  35.11 &  38.84 &  39.50 &  46.06 &  15.68 &  21.29 \\ 
     & GraphLocator & \textbf{75.49} &  \textbf{79.06} &  \textbf{47.34} &  \textbf{53.67} &  \textbf{60.40} &  \textbf{65.38} &  \textbf{42.49} &  \textbf{46.41} &  \textbf{46.32} &  \textbf{54.06} &  \textbf{25.92} &  \textbf{29.36} \\ 
    \hline 
    \multirow{4}{*}{Claude-3.5} & Agentless & 74.24 &  79.19 &  30.92 &  43.03 &  65.41 &  70.40 &  40.65 &  47.47 &  32.50 &  40.19 &  22.94 &  25.73 \\
     & LocAgent & 83.36 &  88.19 &  33.00 &  44.40 &  63.91 &  70.38 &  7.90 &  12.61 &  50.45 &  61.19 &  6.80 &  10.50 \\
     & CoSIL & 67.62 &  71.33 &  18.18 &  27.87 &  59.90 &  64.64 &  38.17 &  44.42 &  45.60 &  53.94 &  20.13 &  26.31 \\
     & GraphLocator & \underline{\textbf{84.97}} &  \underline{\textbf{88.42}} &  \underline{\textbf{48.13}} &  \underline{\textbf{55.73}} &  \underline{\textbf{74.44}} &  \underline{\textbf{80.27}} &  \underline{\textbf{43.68}} &  \underline{\textbf{49.98}} &  \underline{\textbf{60.32}} &  \underline{\textbf{69.39}} &  \underline{\textbf{28.86}} &  \underline{\textbf{33.35}} \\
     \bottomrule
    \rowcolor{mygray}\multicolumn{14}{c}{\textbf{Multi-SWE-bench Java}} \\
    \multirow{3}{*}{GPT-4o} & Agentless & 57.26 &  66.89 &  29.56 &  39.01 &  47.58 &  59.61 &  28.86 &  36.18 &  34.82 &  48.71 &  16.10 &  21.09 \\ 
      & CoSIL & 50.00 &  57.68 &  18.36 &  25.94 &  43.55 &  52.12 &  33.95 &  37.40 &  36.61 &  44.02 &  15.68 &  20.99 \\ 
      & GraphLocator & \textbf{64.52} &  \textbf{71.03} &  \textbf{48.20} &  \textbf{51.73} &  \textbf{54.03} &  \textbf{64.69} &  \textbf{46.40} &  \textbf{47.19} &  \textbf{39.29} &  \textbf{51.51} &  \textbf{21.30} &  \textbf{24.65} \\ 
    \hline 
    \multirow{3}{*}{Claude-3.5} & Agentless & 58.87 &  69.18 &  34.64 &  43.63 &  49.19 &  62.00 &  32.60 &  39.96 &  25.00 &  33.11 &  19.72 &  21.06 \\
      & CoSIL & 58.87 &  65.70 &  20.40 &  29.25 &  45.97 &  55.62 &  32.90 &  37.43 &  31.25 &  42.05 &  17.40 &  21.69 \\
      & GraphLocator &  \underline{\textbf{70.16}} &  \underline{\textbf{76.03}} &  \underline{\textbf{43.26}} &  \underline{\textbf{48.50}} &  \underline{\textbf{58.87}} &  \underline{\textbf{70.44}} &  \underline{\textbf{42.10}} &  \underline{\textbf{46.32}} &  \underline{\textbf{46.43}} &  \underline{\textbf{60.13}} &  \underline{\textbf{24.56}} &  \underline{\textbf{26.64}} \\
    \bottomrule
    \end{tabular}}
    \label{tab:effectiveness}
\end{table*}

\emph{GraphLocator consistently achieves superior performance across datasets and granularities among LLM-based approaches.}
Specifically, on both Python (SWE-bench Lite and LocBench) and Java (Multi-SWE-bench) datasets, \textsc{GraphLocator} achieves the highest localization performance at file-, module-, and function-level compared to Agentless, LocAgent, and CoSIL. Averaged across the three datasets, \textsc{GraphLocator} improves F1 scores on GPT-4o by 21.33\%, 14.18\%, and 8.16\% at file-, module-, and function-level, respectively. On Claude-3.5, the corresponding gains are 16.79\%, 12.64\%, and 9.46\%. 
Since SWE-bench Lite only contains single-file issues, file-level SL and REC are identical. 
Compared with Agentless, \textsc{GraphLocator} substantially increases REC by employing a flexible agentic hierarchical traversal instead of the fixed traversal strategy used in Agentless, allowing it to better adapt to varying granularities in issue descriptions. 
Compared with LocAgent, \textsc{GraphLocator} improves F1 by 14.83–30.68\% across granularities on GPT-4o and 11.33–37.37\% on Claude-3.5, demonstrating higher precision and more coherent causal reasoning.
Compared to CoSIL, which performs graph-based pruning guided solely by contextual relevance, \textsc{GraphLocator} achieves consistently better performance across all granularities. This improvement highlights the limitations of relevance-based heuristics: they can prune away potentially important nodes, accumulate errors or cause under-expansion during multi-step traversal.

\emph{While the embedding-based approach SWERank-Large attains higher SL and REC at coarse granularities (file and module levels), it suffers from reduced effectiveness at finer granularity (function level).} 
Specifically, SWERank-Large exhibits better performance at coarse granularities compared to GPT-4o by retrieving a large number of code snippets. However, at function level, it underperforms both GPT-4o and Claude-3.5 with GraphLocator, and its PRE remains very low (e.g., 6.57\% and 7.88\% PRE on SWE-bench Lite and LocBench). In contrast, \textsc{GraphLocator} not only maintains competitive SL and REC but also substantially improves precision by causal structure discovering and dynamic issue disentangling. Consequently, the overall F1 gains of \textsc{GraphLocator} over SWERank reach up to 15.68-21.32\% at the function level and 10.55-18.21\% at file-level granularities, demonstrating that addressing causality rather than similarity is crucial in code localization.

\emph{Graph-guided causal reasoning significantly improves precision without sacrificing recall.}
\textsc{GraphLocator} consistently boosts PRE compared with both embedding-based and LLM-based approaches. For example, on the Python datasets SWE-bench Lite and LocBench, \textsc{GraphLocator} achieves nearly 4$\times$ the function-level PRE of the LLM-based agent approach LocAgent when using Claude-3.5, which relies on unconstrained agent traversal over the repository graph. This implies that without explicit structural guidance, agent-based methods tend to over-expand the search space, introducing spurious candidates and severely lowering PRE. In contrast, \textsc{GraphLocator} leverages graph-guided abductive reasoning to constrain expansions to causally plausible neighborhoods, thereby filtering irrelevant vertices while still maintaining a high REC.
A similar trend is observed on the Multi-SWE-bench Java, where \textsc{GraphLocator} consistently improves PRE over both Agentless and CoSIL by 4.84-29.84\%. This demonstrates that the benefits of graph-guided causal reasoning generalize beyond Python repositories, enabling more accurate localization where interdependent modules and coordinated changes are more prevalent~\cite{zan2025multiswebench}.

\emph{Effectiveness is more pronounced at finer granularities.}
The function-level improvements are more significant than file- or module-level. This suggests that \textsc{GraphLocator}’s ability to model multi-step causal chains and sub-issue dependencies is particularly beneficial for fine-grained issue localization. At coarser levels, such as file or module, baseline approaches can already achieve relatively high recall because the search space of coarser granularity is smaller and thus less precise reasoning is sufficient. However, at the function level, precise identification of the root cause requires distinguishing between closely related but non-causal code entities, a task where traditional embedding- or agent-based methods often fail due to noise and over-expansion. By explicitly guiding the reasoning process with causal graphs, \textsc{GraphLocator} narrows the search to structurally plausible candidates, enabling significant precision gains without losing recall.

\begin{tcolorbox}[left=2pt,right=2pt,top=0pt,bottom=0pt,boxrule=0.5pt]
\textbf{Answer to RQ1: } 
\textsc{GraphLocator} consistently outperforms baselines in issue localization across Python and Java datasets, achieving absolute improvements of +19.49\% in function-level recall and +11.89\% in precision with Claude-3.5.
\end{tcolorbox}

\subsection{RQ2: Generalizability}

In this subsection, we systematically evaluate the generalizability of issue localization approaches under varying levels of task complexity. 
Specifically, we focus on the task complexity in two key dimensions~\cite{zan2025multiswebench, chen2025locagent}: (1) symptom-to-cause distance and (2) the number of involved functions.
For simplification, all the experiments are conducted based on Claude-3.5.

\begin{figure}[t]
    \begin{minipage}[b]{0.49\linewidth}
        \centering
        \includegraphics[width=\linewidth]{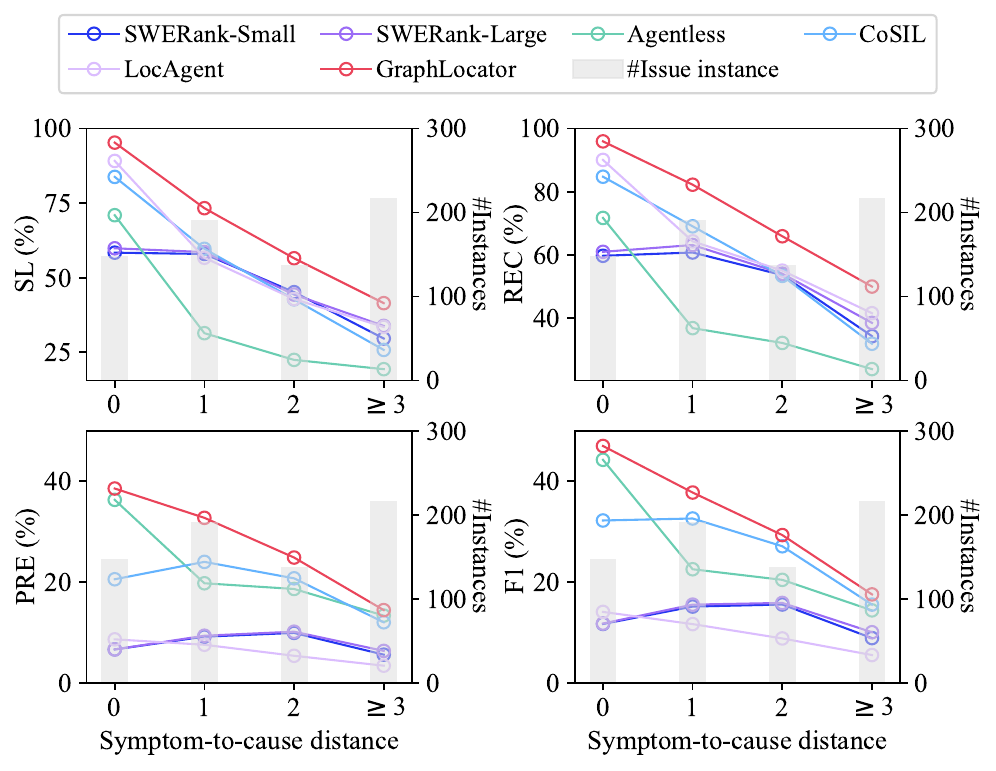}
        \vspace{-3mm}
        \caption{Generalizability across symptom-to-cause distance. 
        Performance decreases as the symptom-to-cause distance grows, while \textsc{GraphLocator} consistently achieves the best results among baselines.}
        \label{fig:symptom_cause_distance}
    \end{minipage}
    \hfill
    \begin{minipage}[b]{0.49\linewidth}
        \centering
        \includegraphics[width=\linewidth]{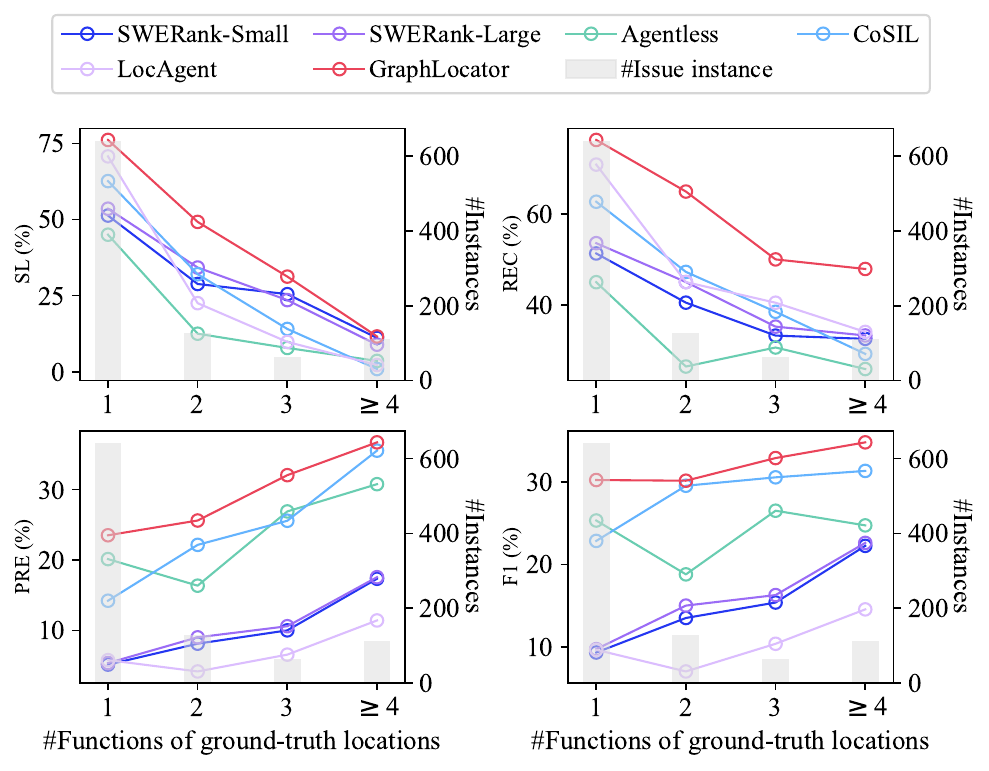}
        \vspace{-3mm}
        \caption{Generalizability across the number of involved functions. 
        As the number of ground-truth functions increases, performance drops for all methods, but \textsc{GraphLocator} remains the most robust.}
        \label{fig:one_to_many}
    \end{minipage}
\end{figure}

\subsubsection{Generalizability on Symptom-to-Cause Distance}
In this subsection, we evaluate the capability of the issue localization approaches to identify root causes given an issue description.
We define the symptom-to-cause distance as the number of traversal hops in the RDFS required to reach all ground-truth locations.
To compute this distance, we first extract keywords from the issue description using Claude-3.5. These keywords are then mapped to RDFS nodes with exactly matching names. Instances without extractable keywords or without valid node mappings are excluded, forming the set $\mathcal{C}$.
In parallel, we map the ground-truth functions to another node set $\mathcal{T}$. The symptom-to-cause distance is then defined as:
\begin{equation*}
    d(\mathcal{T}, \mathcal{C}) = \max_{t \in \mathcal{T}}\min_{c \in \mathcal{C}}d(t, c)
\end{equation*}
where $d(t, c)$ denotes the shortest path distance between node $t$ and node $c$ on the RDFS.
For cases where a ground-truth function is explicitly mentioned in the issue description, the distance is zero.

Fig.~\ref{fig:symptom_cause_distance} shows the generalizability of different approaches with respect to symptom-to-cause distance. For Python-specific approaches (SWERank-Small/Large and LocBench), results are averaged over SWE-bench Lite and LocBench, while other methods are averaged over SWERank-Small/Large, LocBench, and Multi-SWE-bench Java.
From Fig.~\ref{fig:symptom_cause_distance}, it can be observed three key trends:
(1) Performance degradation with distance: SL and REC consistently decline as symptom-to-cause distance increases across all methods.
(2) \textsc{GraphLocator}’s superiority: \textsc{GraphLocator} consistently outperforms baselines. At distance 0, where it achieves 95.59\% REC and the highest precision, validating the effectiveness of its layer-wise RDFS.
(3) Limitations of baselines to balance REC and PRE:
Embedding-based approaches (SWERank-Small/Large) show the largest gap to \textsc{GraphLocator} at distance 0.
Agentless suffers sharp declines due to the lack of structural context, though it maintains relatively high precision by retaining only the top-n most suspicious entities at each layer. Among graph-based agent methods, CoSIL achieves PRE and F1 scores close to \textsc{GraphLocator} at distances 3 and 4, benefiting from its pruner module that filters out some false positives; however, the absence of explicit causal-path reasoning results in low REC and SL. In contrast, LocAgent exhibits particularly low PRE and F1, as it fails to distinguish between relevant and true causal code, retrieving a large number of non-causal snippets.

\subsubsection{Generalizability on Number of Involved Functions}
Fig.~\ref{fig:one_to_many} shows the generalizability of different approaches with respect to the number of functions in ground-truth. Similar to Fig.~\ref{fig:symptom_cause_distance}, for Python-specific approaches, results are averaged over SWE-bench Lite and LocBench, while others are averaged over both the Python and Java datasets.

From Fig.~\ref{fig:one_to_many}, we observe that among all approaches, \textsc{GraphLocator} achieves the best balance between REC and PRE across varying numbers of functions in ground truth.
Compared with embedding-based approaches (SWERank-Small/Large), \textsc{GraphLocator} maintains substantially higher SL and REC, highlighting its robustness in complex multi-function cases.
Agentless, lacking structural reasoning, shows sharp declines in SL and REC as the number of functions grows.
LocAgent performs comparably to \textsc{GraphLocator} when only a single function is involved; however, as the number of functions increases, it loses the ability to trace multi-function dependencies due to the absence of CIG guidance, which is leveraged by \textsc{GraphLocator}.
CoSIL benefits from its pruning strategy, yielding higher precision but at the cost of reduced recall, while LocAgent further suffers from poor discrimination of true causal functions, leading to consistently low PRE and F1.
Additionally, there is an opposite trend between SL/REC and PRE/F1: as the number of ground-truth functions increases, SL and REC steadily decline, while PRE and F1 improve.
This is partially because a larger number of ground-truth functions expands the search space, making it more difficult to recover all functions (lowering SL and REC), yet simultaneously increasing the probability that predictions overlap with part of the correct functions (raising PRE and F1).

\begin{tcolorbox}[left=2pt,right=2pt,top=0pt,bottom=0pt,boxrule=0.5pt]
\textbf{Answer to RQ2: }
As both the symptom-to-cause distance and the number of ground-truth functions increase, \textsc{GraphLocator} consistently outperforms baseline approaches and demonstrates superior effectiveness in balancing recall and precision.
\end{tcolorbox}

\subsection{RQ3: Ablation Study}
In this subsection, we systematically evaluate the contributions of key components in \textsc{GraphLocator}.  
For Phase I (symptom vertex locating), we ablate the tools \textit{search\_vertex} and \textit{search\_edge} individually. For Phase II (dynamic CIG discovery), two variants are considered: (i) \emph{w/o priority queue}, which replaces the priority queue with a simple first-in-first-out (FIFO) queue; and (ii) \emph{w/o CIG guidance}, which excludes the CIG from the prompt, retaining only the already selected code entities and removing causal-structure guidance. All experiments use Claude-3.5, and results are averaged across SWE-bench Lite, LocBench, and Multi-SWE-bench Java.

\begin{table*}[t]
    \centering
    \caption{Ablation study of components in \textsc{GraphLocator}. Results are averaged across the 987 issues of the three datasets (i.e., SWE-bench Lite, LocBench, and Multi-SWE-bench Java) based on Claude-3.5.}
    \vspace{-3mm}
     \scalebox{0.78}{
    \begin{tabular}{lcccc|cccc|cccc}
    \toprule
    \multirow{2}{*}{\textbf{Approach}} & \multicolumn{4}{c|}{\textbf{File-level}} & \multicolumn{4}{c|}{\textbf{Module-level}} & \multicolumn{4}{c}{\textbf{Function-level}}  \\
    \cmidrule{2-13}
    & \textbf{SL} & \textbf{REC} & \textbf{PRE} & \textbf{F1} & \textbf{SL} & \textbf{REC} & \textbf{PRE} & \textbf{F1} & \textbf{SL} & \textbf{REC} & \textbf{PRE} & \textbf{F1} \\
    \hline
    \textbf{GraphLocator} & \textbf{84.44} &  \textbf{87.14} &  \textbf{46.81} &  \textbf{54.65} &  \textbf{73.66} &  \textbf{79.09} &  \textbf{41.90} &  \textbf{48.42} &  \textbf{61.73} &  \textbf{69.41} &  \textbf{25.91} &  \textbf{30.91} \\
    \quad w/o search\_vertex & 35.40 &  37.23 &  24.56 &  27.04 &  38.90 &  41.78 &  26.96 &  29.49 &  25.48 &  29.32 &  12.29 &  14.23 \\
    \quad w/o search\_edge & 74.36 &  78.44 &  42.04 &  49.34 &  64.04 &  69.68 &  37.72 &  43.40 &  51.59 &  58.77 &  23.85 &  28.45 \\
    \quad w/o priority queue & 75.08 &  78.98 &  44.00 &  51.10 &  64.97 &  70.17 &  40.91 &  46.17 &  54.76 &  62.27 &  24.12 &  29.54 \\
    \quad w/o CIG guided & 75.89 &  80.09 &  33.83 &  41.77 &  68.05 &  73.99 &  32.11 &  39.24 &  59.30 &  66.50 &  16.40 &  21.50\\
    \bottomrule
    \end{tabular}}
    \label{tab:ablation-study}
\end{table*}

Tab.~\ref{tab:ablation-study} reports the performance impact of ablating each component in \textsc{GraphLocator}.
The localization performance of \textsc{GraphLocator} suffers a significant drop after the removal of each of its component.
Specifically, in the symptom vertices locating phase, \textit{search\_vertex} serves as the primary bridge linking issue descriptions to RDFS vertices; its removal collapses function-level F1 from 30.91\% to 14.23\%. \textit{search\_edge} provides supplementary context, and its absence also degrades performance, highlighting the necessity of both tools for accurate seed identification.  
In the dynamic CIG discovering phase, CIG-based guidance in prompts ensures that the LLM maintains awareness of previously identified causal dependencies, which is critical for the precise of causal reasoning.
Priority-driven graph expansion, on the other hand, improves recall by steering reasoning toward the most promising sub-issues.
This result further implies that though the LLM-estimated causal dependency probabilities may not exactly reflect the true values, they provide a useful relative measure of the importance among sub-issues.

\begin{tcolorbox}[left=2pt,right=2pt,top=0pt,bottom=0pt,boxrule=0.5pt]
\textbf{Answer to RQ3: }
The performance of \textsc{GraphLocator} degrades after
the removal of each of its key components. Each component, from search tools to priority-driven graph expansion, and graph-guided reasoning, directly enables effective localization.
\end{tcolorbox}

\subsection{RQ4: Cost}

\begin{figure*}[t]
    \centering
    \begin{minipage}{0.58\linewidth}
        \centering
        \includegraphics[width=1\linewidth]{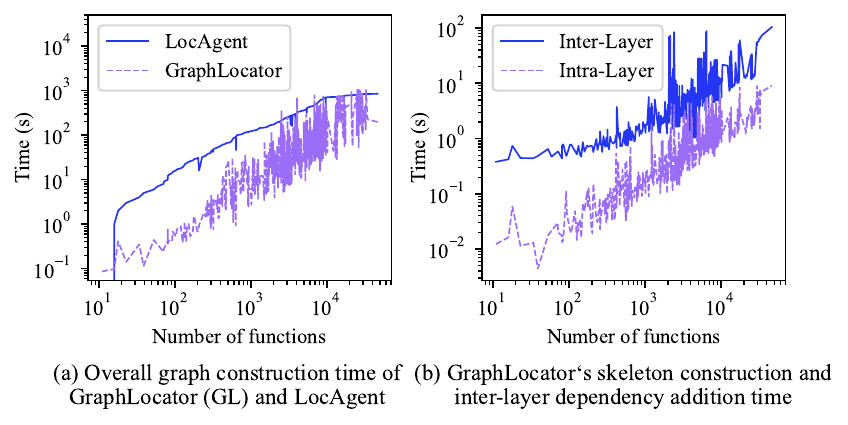}
        \vspace{-8mm}
        \caption{Time efficiency of GraphLocator and LocAgent.}
        \label{fig:time}
    \end{minipage}
    \hspace{-2mm}
    \begin{minipage}{0.4\linewidth}
        \centering
        \captionof{table}{Cost for interaction with LLM}
        \vspace{-3mm}
        \label{tab:cost}
        {\renewcommand{\arraystretch}{1.1}
        \scalebox{0.7}{
        \begin{tabular}{lccc}
        \toprule
         \textbf{Approach} & \textbf{Input (k)} & \textbf{Output (k)} & \textbf{Cost (\$)} \\
         \hline
        \rowcolor{mygray}\multicolumn{4}{c}{\textbf{GPT-4o-2024-1120}} \\
         Agentless    &  26.51 & 0.60 & 0.13\\
         LocAgent  & 211.23 & 1.74 & 1.08 \\
         CoSIL & 12.93 & 0.86 & 0.08 \\
         GraphLocator & 99.44 & 7.56 & 0.61 \\
         \hline
         \rowcolor{mygray}\multicolumn{4}{c}{\textbf{Claude-3.5-Sonnet-20241022}} \\
         Agentless    & 62.56 & 0.95 & 0.19 \\
         LocAgent    & 231.68 & 2.85 & 0.74 \\
         CoSIL & 14.24 & 0.91 & 0.06 \\
         GraphLocator & 156.80 & 6.71 & 0.57\\
         \bottomrule
        \end{tabular}}}
    \end{minipage}
\end{figure*}

\subsubsection{Graph Construction Time Efficiency.}
In Fig.~\ref{fig:time}, we study the time efficiency of \textsc{GraphLocator}'s graph construction time.
The graph construction time of \textsc{GraphLocator} consists of two parts: the inter-layer skeleton construction time before agent processing the issue and the intra-layer dependencies lazy loading during the agent process.
In Fig.~\ref{fig:time}(a), we compare the overall time of \textsc{GraphLocator} and LocAgent. 
In Fig.~\ref{fig:time}(b), we study how the inter-layer skeleton and intra-layer dependencies construction time scale as the number of functions in a repository grows.
We do not compare with the graph-based approach CoSIL since its graph is dynamically constructed by prompting LLM instead of static analysis, thus is hard to track.

As shown in Fig.~\ref{fig:time}(a), \textsc{GraphLocator} achieves higher average time efficiency in graph construction when the repository contains more than 20 functions.
The RDFS used by \textsc{GraphLocator} is more fine-grained and complicate than that of LocAgent since it also incorporate entities such as class field and global variables, intuitively its fully construction is more time-consuming.
However, the lazy loading mechanism used by \textsc{GraphLocator} effectively reduce the time by only load the dependencies that is useful instead of the fully construction used by \textsc{GraphLocator}.
The fluctuation of \textsc{GraphLocator} on overall graph is resulting by the number of useful dependencies for resolving different issue instance varied.
In Fig.~\ref{fig:time} (b), we look into how the inter-layer skeleton construction time and intra-layer dependencies construction time scale as the number of functions grows.
Both times grow with the number of functions. For a repository with 10k functions, intra-layer skeleton construction takes under 20 seconds, while inter-layer static analysis of a single function ranges from 0.5 to 5 seconds. Time efficiency can be further improved by parallelizing the adding of intra-layer dependencies.

\subsubsection{Token Consumption}
Tab.~\ref{tab:cost} presents the token consumption and associated cost of different approaches across two representative LLMs. Compared with LocAgent, \textsc{GraphLocator} achieves a substantial reduction in interaction overhead. Under GPT-4o, \textsc{GraphLocator} requires less than half of LocAgent’s input tokens (99.44k vs. 211.23k), reducing token usage by 52.9\% and cost by 43.5\%. A similar trend holds under Claude-3.5, where \textsc{GraphLocator} reduces input tokens by 32.3\% and cost by 22.9\% relative to LocAgent. While \textsc{GraphLocator} consumes more tokens than Agentless and CoSIL due to its repository-structural reasoning, the additional cost remains moderate. Overall, \textsc{GraphLocator} strikes a favorable trade-off between accuracy and efficiency, offering significantly lower token consumption than agent-based baselines while maintaining strong localization performance.

\begin{tcolorbox}[left=2pt,right=2pt,top=0pt,bottom=0pt,boxrule=0.5pt]
\textbf{Answer to RQ4: }
\textsc{GraphLocator} demonstrates efficient graph construction and substantially lower token consumption compared with agent-based baseline LocAgent, achieving strong localization performance with acceptable overall cost.
\end{tcolorbox}

\section{Discussion}

\subsection{Impact on Downstream Issue Resolving}

In this subsection, we examine how issue localization performance influences downstream issue resolving. 
We select two representative issue resolving frameworks: the procedural Agentless~\cite{xia2024agentless} and the recent agent-based Trae Agent~\cite{traeresearchteam2025traeagent}. Prior studies~\cite{jiang2025cosil, xia2024agentless, yu2025orcaloca} have suggested a positive correlation between file- and function-level success localization (SL) rate and resolving performance; here, we systematically quantify this relationship.

For integrating localization approaches into Agentless, we replace the original location module of Agentless by baseline localization approaches and GraphLocator. 
For Trae Agent, we serialize the localization results, including file paths and corresponding code snippets, in the order produced by each baseline and concatenate this information with the original issue description. To evaluate the contribution of \textsc{GraphLocator}'s causal issue graph (CIG) structure, we consider two settings:
(1) GraphLocator (w/o struct.): code entities are serialized in topological order and appended to the issue description, without representing structural dependencies.
(2) GraphLocator (w/ struct.): the generated CIG is serialized in topological order and represented using the Mermaid format, preserving structural relations.
Both frameworks leveraging multi-sampling mechanism to improve issue resolved rate; for this evaluation, we use a single-sample setting (num\_sampling=1) with greedy decoding for proof of concept and budget saving reduce computational cost. Evaluation is performed on SWE-bench Lite and Multi-SWE-bench Java, excluding LocBench due to missing unit tests information for correctness validation. The results are shown in Tab.~\ref{tab:resolve_rate}.

\begin{table}[h]
    \centering
    \caption{Resolved rate on SWE-bench Lite and Multi-SWE-bench Java using Claude-3.5 with a single-sample evaluation. \textbf{GraphLocator (GL)} is abbreviated as GL. File SL and Func SL represent file- and function-level success location rate, respectively. Improv. indicates the relative improvement in resolved rate over downstream framework. All results are in percentage (\%).}
    \label{tab:resolve_rate}
    \scalebox{0.7}{
    \begin{tabular}{lcccc|lcccc}
    \toprule
     \textbf{Approach} & \textbf{File SL} & \textbf{Func SL} & \textbf{Resolved} & \textbf{Improv.} & \textbf{Approach} & \textbf{File SL} & \textbf{Func SL} & \textbf{Resolved} & \textbf{Improv.}\\
    \hline
    \rowcolor{mygray}\multicolumn{10}{c}{\textbf{SWE-bench Lite}} \\
    \textbf{Agentless}  & 80.00 & 37.91 & 25.33 & - & \textbf{Trae Agent} & 58.33 & 40.07 & 25.00 & -\\
    \quad +SWERank-Large & 84.33 & 55.23 & 26.33 & $\uparrow$ 3.95 & \quad +SWERank-Large & 84.33 & 55.23 & 28.33 & $\uparrow$ 13.32 \\
    \quad +LocAgent & 84.33 & 58.48 & 25.67 & $\uparrow$ 1.34 & \quad +LocAgent & 84.33 & 58.48 & 27.00 & $\uparrow$ 8.00 \\
    \quad +CoSIL & 82.33 & 59.21 & 26.00 & $\uparrow$ 2.65 & \quad +CoSIL & 82.33 & 59.21 & 28.67 & $\uparrow$ 14.68 \\
    \quad +GL (w/o struct.) & 89.33 & 70.76 & 26.33 & $\uparrow$ 3.95 & \quad +GL (w/o struct.) & 89.33 & 70.76 & 29.00 & $\uparrow$ 16.00 \\
    \quad \textbf{+GL (w/ struct.) } & \textbf{89.33} & \textbf{70.76} & \textbf{28.67} & \textbf{$\uparrow$ 13.19 }& \quad \textbf{+GL (w/ struct.)} & \textbf{89.33} & \textbf{70.76} & \textbf{30.67} & \textbf{$\uparrow$ 22.68} \\
    \hline
    \rowcolor{mygray}\multicolumn{10}{c}{\textbf{Multi-SWE-bench Java}} \\
    \textbf{Agentless}  & 58.57 & 25.00 & 8.59 & - & \textbf{Trae Agent}  & 51.61 & 33.04 & 20.31 & - \\
    \quad +CoSIL  & 58.87 & 31.25 & 12.50 & $\uparrow$ 45.54 & \quad +CoSIL & 58.87 & 31.25 & 14.06 & $\downarrow$ 30.77 \\
    \quad +GL (w/o struct.)  & 70.16 & 46.43 & 13.28 & $\uparrow$ 54.60 & \quad +GL (w/o struct.) & 70.16 & 46.43 & 22.65 & $\uparrow$ 11.52\\
    \quad \textbf{+GL (w/ struct.)} & \textbf{70.16} & \textbf{46.43} & \textbf{14.06} & \textbf{$\uparrow$ 63.68} & \quad \textbf{+GL (w/ struct.)} & \textbf{70.16} & \textbf{46.43} & \textbf{23.44} & \textbf{$\uparrow$ 15.41}\\
    \bottomrule
    \end{tabular}}
\end{table}

From Tab.~\ref{tab:resolve_rate}, several key observations can be drawn:
(1) Improved localization generally leads to higher resolved rates: across both frameworks, integrating \textsc{GraphLocator} consistently increases resolved rate compared to baselines. For example, on SWE-bench Lite, \textsc{GraphLocator} (w/ struct.) improves Trae Agent’s resolved rate from 25.00\% to 30.67\%.
However, there is also an exception: on SWE-bench Lite, +SWERank-Large achieves a higher resolved rate than LocAgent despite lower function-level SL, likely due to LocAgent’s lower precision and F1 (Tab.\ref{tab:effectiveness})
(2) Structural and explainable context provided by the Mermaid-serialized CIG, consistently enhances resolved rates, indicating that explicit encoding of causal and relationships helps LLMs better reason about necessary code changes.
(3) Trae Agent benefits more from high-quality localization than Agentless, indicating that an agentic workflow can more effectively leverage prompt-based cues to guide autonomous issue resolving.
(4) While accurate localization is undeniably crucial, how to effectively utilizing this information to generate correct patches remains a significant challenge. The absolute resolved rates, though improved, indicate that translating accurate localization into correct fixes involves additional complexities beyond mere identification of code to be modified.
Knowing \emph{where} to change is not enough; it is also essential to know \emph{how} to change it. Therefore, more sophisticated mechanisms for generating fix patches are required in future work to bridge this gap and fully leverage the potential of structural code understanding.

\subsection{Threats to Validity}
\emph{Internal validity.}
The internal threats to validity lie in the implementation of baseline approaches, the potential data leakage in LLMs, and the definition of ground-truth location.
For the implementation of baselines, we directly use official implementations released on GitHub and configure them according to their default settings to ensure fair comparison.
For the adaption of CoSIL to Java version, we implement by directly use the Magentless realized by Multi-SWE-bench~\cite{zan2025multiswebench}.
For adapting CoSIL to Java, we follow the implementation of Magentless provided in Multi-SWE-bench~\cite{zan2025multiswebench} to generate the repository structure required by CoSIL.
Regarding LLMs, while more recent models could further validate \textsc{GraphLocator}, they pose a higher risk of data leakage. 
To mitigate this, we avoid very recent releases (e.g., Claude-3.7-Sonnet, Claude-4-Opus, GPT-5) and instead select two widely adopted LLMs whose training data are less likely to overlap with LocBench, which was collected after October 2024. 
Moreover, as shown in Tab.~\ref{tab:effectiveness}, \textsc{GraphLocator} consistently outperforms baselines even under potential leakage, suggesting that its improvements are not merely a result of memorization.
Finally, issue localization does not always have a single definitive solution, as alternative code changes may also correctly resolve a given issue. Following prior work~\cite{chen2025locagent, reddy2025swerank, jiang2025cosil}, we use human-written fix patches to extract ground-truth locations, which reflect commonly accepted fixes while acknowledging that alternative valid solutions may exist.

\emph{External validity.}
The primary external threat lies in the generalizability of our findings beyond the current evaluation setting, including applicability to other programming languages and to diverse types of repositories.
Our evaluations focus on open-source repositories in two mainstream programming languages: Python and Java, so our results are limited in this scope. 
Nevertheless, \textsc{GraphLocator} can be extended to additional languages by constructing the RDFS from abstract syntax trees and reusing the same reasoning framework with relatively modest effort.
Regarding repository diversity, our evaluation includes 164 repositories from LocBench, 12 from SWE-bench Lite, and 9 from Multi-SWE-bench Java, covering a broad range of domains and sizes.
Although not exhaustive, this selection provides a representative basis for assessing the effectiveness of \textsc{GraphLocator} across different settings.

\section{Related Work}

\paragraph{Issue Localization Approaches.}
Issue localization aims to map natural language descriptions of issues to the corresponding code entities that require modification. Existing approaches can be largely categorized into \emph{embedding-based} and \emph{LLM-based} approaches.
\emph{Embedding-based methods} formulate the task as semantic matching between issue descriptions and code entities. Representative models such as CodeSage~\cite{codesage}, CodeRankEmbed~\cite{sureshcornstack}, and SWERankEmbed~\cite{reddy2025swerank} learn joint embedding spaces for natural language and code, ranking candidate entities by similarity. While lightweight, these methods require complex indexing infrastructures and incur storage and maintenance overheads. More importantly, they overlook structural and dependency relations within repositories, limiting their ability to trace true root causes~\cite{chen2025locagent} or capture interdependent changes.
\emph{LLM-based approaches} leverage the reasoning and code understanding capabilities of LLMs~\cite{zheng2025towards, haroon2025accurately, chen2025towards, plaat2024reasoning}. These approaches can be further divided into \emph{procedural} and \emph{agentic} workflows. 
The procedural workflow follows fixed hierarchical steps guided by prompts. For example, Agentless~\cite{xia2024agentless} localizes issues progressively from files to classes to functions, while BugCerberus~\cite{chang2025bridging} employs specialized LLMs trained for different levels. Although structured, such approaches neglect code structure and cannot backtrack once errors occur.
The agentic workflow treats LLMs as autonomous agents capable of tool use and exploration. Examples include SWE-Agent~\cite{yang2024swe}, OpenHands~\cite{wang2024openhands}, and Trae Agent~\cite{traeresearchteam2025traeagent} and MoatlessTools~\cite{orwall2024moatless}. To improve structural awareness, graph-based variants have been proposed~\cite{ma2024understand, ouyang2024repograph, yu2025orcaloca,jiang2025cosil,chen2025locagent}.
For example, OrcaLoca~\cite{yu2025orcaloca} models hierarchical and call relations using relevance-priority scheduling.
LocAgent~\cite{chen2025locagent} parses repositories into heterogeneous graphs, enabling tool-assisted exploration. CoSIL~\cite{jiang2025cosil} extends this by pruning traversal paths using contextual relevance to reduce noise.
While agentic graph-based approaches enhance structural awareness, they still rely on superficial relevance, lack explicit modeling of causality, and fail to effectively capture interdependent code entities.

\emph{Causal Reasoning Ability of LLMs.}
Causal reasoning is a fundamental aspect of human cognition and is considered essential for advancing machine intelligence~\cite{pearl2019seven}. Traditional frameworks, such as structural causal models (SCMs)\cite{pearl2009causality} and the potential outcome framework\cite{imbens2015causal}, provide systematic definitions and methods for discovering causal relationships~\cite{spirtes2016causal, nogueira2022methods, vowels2022d} and estimating causal effects~\cite{winship1999estimation, yao2021survey}. However, these approaches primarily target tabular data and are less effective for reasoning over natural language.
Recent advancements in LLMs offer new opportunities to extend causal inference to textual and code contexts~\cite{feder2022causal, ma2024causal}. Causal reasoning tasks can be grouped into three categories: causal discovery (recovering latent causal structures), cause attribution (identifying potential causes), and causal effect estimation (measuring the impact of causal variables)\cite{sun2023survey, kiciman2023causal, wu2024causality, jin2023can, zhiheng2022can}. While LLMs show promise in uncovering causal relationships\cite{liu2025large, ashwani2024cause, long2023can, zhou2024causalbench}, they remain limited in counterfactual reasoning, which requires evaluating hypothetical scenarios~\cite{sun2023survey, li2023counterfactual, wu2024reasoning, chi2024unveiling}.
In this paper, \textsc{GraphLocator} leverages the causal discovery capabilities of LLMs to infer latent causal structures from observed code snippets, enabling more precise identification of interdependent sub-issues.

\section{Conclusion}
In this paper, we present \textsc{GraphLocator}, an LLM-based approach for issue localization that addresses the fundamental challenges of symptom–to-cause and one-to-many mismatches between issue description and source code. By constructing a causal issue graph (CIG), \textsc{GraphLocator} captures multi-hop causal relationships among sub-issues and their corresponding code entities, enabling iterative abductive reasoning and dynamic disentangling of interdependent components. Extensive experiments on three real-world Python and Java datasets demonstrate that \textsc{GraphLocator} consistently outperforms existing baselines, achieving substantial improvement in function-level localization recall and precision, as well as improved performance in both symptom–to-cause and one-to-many scenarios. These results highlight the effectiveness of leveraging causal structure discovery and dynamic issue
disentangling to bridge the semantic gap between natural language issues and source code, offering a promising direction for more accurate and interpretable automated issue localization.

\section*{Data Availability}
The prompt templates and code of \textsc{GraphLocator} are released at \url{https://github.com/oceaneLIU/GraphLocator}.
For baselines, we use their publicly available source code released on GitHub, including SWERank (\url{https://github.com/SalesforceAIResearch/SweRank}), CoSIL (\url{https://github.com/ZhonghaoJiang/CoSIL}), LocAgent (\url{https://github.com/gersteinlab/LocAgent}), Agentless (\url{https://github.com/OpenAutoCoder/Agentless}), and Trae Agent(\url{https://github.com/bytedance/trae-agent}).

All datasets used in this study are publicly accessible. SWE-bench Lite is available at \url{https://huggingface.co/datasets/SWE-bench/SWE-bench_Lite}, Multi-SWE-bench at \url{https://huggingface.co/datasets/ByteDance-Seed/Multi-SWE-bench}, and LocBench at \url{https://huggingface.co/datasets/czlll/Loc-Bench_V1}. 

\bibliographystyle{ACM-Reference-Format}
\bibliography{sample-sigconf}

\end{document}